\documentclass[aps,twocolumn]{revtex4}

\usepackage{dcolumn}
\usepackage{bm}
\usepackage{amsmath}
\usepackage{graphicx}
\usepackage{amsmath,amssymb}

\newcommand{\pro}[2]{\langle{#1}|{#2}\rangle}
\newcommand{\bra}[1]{\langle{#1}|}
\newcommand{\ket}[1]{|{#1}\rangle}

\begin{document}

\title{On the natural gradient for variational quantum eigensolver}
\author{Naoki Yamamoto}
\affiliation{Department of Applied Physics and Physico-Informatics \& Quantum Computing Center, 
Keio University, Hiyoshi 3-14-1, Kohoku, Yokohama, 223-8522, Japan}
\date{\today}

\begin{abstract}
The variational quantum eigensolver is a hybrid algorithm composed of quantum state driving 
and classical parameter optimization, for finding the ground state of a given Hamiltonian. 
The natural gradient method is an optimization method taking into account the geometric 
structure of the parameter space. 
Very recently,  Stokes et al. developed the general method for employing the natural gradient 
for the variational quantum eigensolver. 
This paper gives some simple case-studies of this optimization method, to see in detail how 
the natural gradient optimizer makes use of the geometric property to change and improve 
the ordinary gradient method. 
\end{abstract}

\maketitle


\section {Introduction}

The field of quantum computing is still far away from the stage where ideal fault-tolerant 
systems are available. 
Hence a recent trend is to seek the potential of quantum-classical hybrid algorithms which 
might have some advantages over purely classical computers. 
The variational method is one such approach; particularly the {\it variational quantum 
eigensolver (VQE)} 
\cite{O'Brien 2014,McClean 2016,Kandala 2017,Temme 2017,AdptVQE 2019,Fujii 2019} 
is a hybrid algorithm using a parametrized quantum computer to drive the quantum 
state and a classical computer for optimizing those parameters, for finding the ground 
state of a given Hamiltonian.

It is clear that the performance of VQE heavily depends on the classical optimization part. 
In fact various types of optimizers have been tested, such as the ordinary gradient and 
simultaneous perturbation stochastic approximation \cite{SPSA}. 
On the other hand, the so-called {\it natural gradient} optimization method \cite{Amari 1998} 
is often used in the classical literature, particularly for machine learning problems. 
Actually the natural gradient is the optimizer that takes into account the geometric structure 
of the parameter space, which is usually very complicated, e.g., in the case of large neural 
networks, and hence it often works very well without being trapped in local minimums or 
plateaus in the parameter space.

Interestingly, it is known in the literature (see e.g., \cite{Carleo 2017,Cirac 2018}) that, 
in the case when the quantum Fubini-Study metric is taken to measure the geometric 
structure, the VQE algorithm with the natural gradient is equivalent to the {\it stochastic 
reconfiguration} \cite{Sorella 2001,Sorella 2004} and the {\it imaginary time evolution (ITE)} 
\cite{Benjamin 2018a,Benjamin 2018b} (a certain correction term might be necessary). 
With this background, very recently, Stokes et al. \cite{Carleo 2019} developed the 
general framework for applying the natural gradient for variational quantum problems 
and demonstrated that it actually works for a particular VQE problem better than some 
standard optimizers; 
in particular, they gave an efficient algorithm for computing the Fubini-Study metric in 
each iteration step of the VQE procedure.

The aim of this paper is, with detailed investigations for some simple VQE problems, 
to show how the natural gradient makes use of the geometric property of the parametrized 
quantum state to realize a better optimization process of the parameters, and accordingly 
to give a suggestion about in what situation the natural gradient method should be used in 
VQE problems.


\section{VQE, Natural gradient, and ITE}

\begin{figure}[htbp]
\includegraphics[width=8cm]{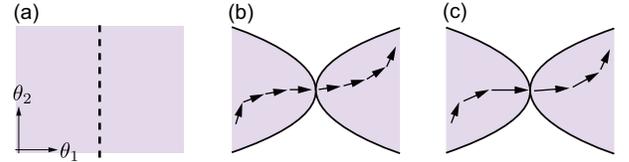}
\caption{
Idea of natural gradient. 
(a) Parameter space; the dotted line represents the set of parameters where 
$f(\theta_1, \theta_2)$ takes the same value for all $\theta_2$. 
(b) Trajectory of the parameters with Euclidean metric. 
(c) Trajectory of the parameters with non-Euclidean metric. 
}
\label{Grad Flow image}
\end{figure}

The basic procedure of VQE is, given a (short) quantum circuit $U(\theta)$ with parameters 
$\theta=(\theta_1, \ldots, \theta_m)$, to repeatedly update $\theta$ so that the mean 
energy $f(\theta)=\bra{\phi(\theta)}H\ket{\phi(\theta)}$ decreases toward its minimum 
for the ansatz $\ket{\phi(\theta)}=U(\theta)\ket{0}$ with $\ket{0}$ the initial state; 
hence VQE is a hybrid algorithm composed of the quantum computing, which generates 
a quantum state possibly hard to classically simulate, and the classical computing, which 
iterates the optimization process of the parameters. 
A simple optimizer is the ordinary gradient descent: 
\begin{equation}
\label{standard gradient}
      \theta_{k+1} = \theta_k - \eta_k \frac{\partial f(\theta)}{\partial \theta}, 
\end{equation}
where $\theta_k$ is the parameter at the $k$th step, $\eta_k$ is the coefficient, and 
$\partial f(\theta)/\partial \theta$ is the gradient vector of $f(\theta)$.

Now note that the dynamics \eqref{standard gradient} assumes that the parameter space is 
a flat Euclidean space. 
However, as will be demonstrated, the actual parameter space is not flat in the sense of 
indistinguishability. 
For instance, for the two-parameters case $f(\theta)=f(\theta_1, \theta_2)$, there might be 
a point $\theta_1=\bar{\theta}_1$ such that $f(\bar{\theta}_1, \theta_2)$ takes the same 
value for all $\theta_2$; 
in this case, because any two different $\theta_2$ cannot be distinguished, this 1-dimensional 
set of parameters (the dotted line in Fig.~\ref{Grad Flow image}(a)) must be regarded as 
a singular point in the parameter space. 
Consequently, the parameter space must be a non-Euclidean one as shown in 
Fig.~\ref{Grad Flow image}(b). 
Around the singular point, roughly speaking the volume of metric becomes small, and thus 
the dynamics of parameters should be modified, taking into account this geometry; 
the natural gradient does this, by stretching the gradient vector as shown in 
Fig.~\ref{Grad Flow image}(c) while the ordinary one \eqref{standard gradient} does not. 
As consequence, the natural gradient drives the parameter point faster than the ordinary 
one, especially around the singular points. 
Mathematically, the natural gradient optimizer updates the parameters according to 
\begin{equation}
\label{natural gradient}
      \theta_{k+1} = \theta_k - \eta_k F(\theta)^{-1}\frac{\partial f(\theta)}{\partial \theta}, 
\end{equation}
where $F(\theta)$ is the Riemannian metric matrix at $\theta$. 
In fact, $F^{-1}\partial f/\partial \theta$ is the steepest descent vector in this Riemanean 
space.

In the VQE setting, the metric can be induced from the indistinguishability of the function 
$f(\theta)=\bra{\phi(\theta)}H\ket{\phi(\theta)}$. 
The most ``detailed" distinguishing way is to measure the distance in the space of pure 
quantum states; 
actually if $\ket{\phi(\theta_A)}=\ket{\phi(\theta_B)}$, then $f(\theta_A)=f(\theta_B)$. 
Typically the Fubini-Study distance is used for this purpose, the infinitesimal version of which 
is given by 
\[
    {\rm Dist}_Q\Big(\ket{\phi(\theta)}, ~ \ket{\phi(\theta+d\theta)} \Big)^2
      = \sum_{i,j}F_{ij}(\theta)d\theta_i d\theta_j, 
\]
where $F=(F_{ij})$ is the Fubini-Study (or more generally the quantum Fisher information) metric: 
\begin{equation}
\label{FS metric}
     F_{ij} = {\rm Re}(\pro{\partial_i \phi}{\partial_j \phi}) 
                 - \pro{\partial_i \phi}{\phi}\pro{\phi}{\partial_j \phi}. 
\end{equation}
Here $\ket{\partial_i \phi}=\partial\ket{\phi(\theta)}/\partial\theta_i$ denotes the 
partial derivative of $\ket{\phi(\theta)}$ with respect to $\theta_i$. 
The singular point is now clearly characterized by the point where the matrix $F$ is not 
of full-rank.

We can introduce another measure for indistinguishability; 
the energy function is now represented in terms of the classical probability distribution 
$p(\theta)=\{p_i(\theta)\}$ as 
\[
    f(\theta) = \sum_i h_i p_i(\theta), ~~
     p_i(\theta) = \bra{\phi(\theta)} E_i \ket{\phi(\theta)}, 
\]
where $\lambda_i$ and $E_i$ are the eigenvalue and the corresponding projection operator 
of $H$, respectively. 
Clearly, if $p(\theta_A)=p(\theta_B)$, then $f(\theta_A)=f(\theta_B)$. 
Typically the distance of probability distributions is measured by the Kullback-Leibler 
divergence, the infinitesimal version of which is given by 
\[
    {\rm Dist}_C\Big( p(\theta), ~ p(\theta+d\theta) \Big)^2
      = \frac{1}{4}\sum_{i,j}F^C_{ij}(\theta)d\theta_i d\theta_j, 
\]
where $F^C=(F^C_{ij})$ is the Fisher information metric: 
\begin{equation}
\label{C Fisher metric}
    F^C_{ij} = {\mathbb E}\Big[ 
                      \Big(\frac{\partial \log p(\theta)}{\partial \theta_i}\Big)
                      \Big(\frac{\partial \log p(\theta)}{\partial \theta_j}\Big) \Big]. 
\end{equation}
As in Ref.~\cite{Carleo 2019}, this paper studies the natural gradient with the quantum 
metric \eqref{FS metric}; 
using Eq.~\eqref{C Fisher metric} is an interesting direction, but only a small comment will 
be given in the next section.

Lastly let us review the theory of ITE \cite{Benjamin 2018a}. 
The basic idea is to project the energy-decreasing (hence non-Unitary) dynamics 
$d\ket{\psi}/dt=-(H-\bra{\psi}H\ket{\psi})\ket{\psi}$ to the ansatz space; 
in the discrete time representation the following parameter update rule governs the 
projected dynamics: 
\begin{equation}
\label{ITE}
      \theta_{k+1} = \theta_k - \eta_k A(\theta)^{-1}\frac{\partial f(\theta)}{\partial \theta}, 
\end{equation}
where $A_{ij} = {\rm Re}(\pro{\partial_i \phi}{\partial_j \phi})$. 
Therefore by adding the second term in Eq.~\eqref{FS metric} to $A_{ij}$, we find that 
the ITE is equivalent to the natural gradient. 
Although a different projection method leads to the ansatz dynamics of ITE which is 
completely equivalent to the natural gradient, as shown in \cite{Benjamin 2018b,Carleo 2019}, 
in this paper let us call Eq.~\eqref{ITE} the ITE. 
Also note that the residual matrix 
$(A-F)_{ij} = \pro{\partial_i \phi}{\phi}\pro{\phi}{\partial_j \phi}$ is positive semidefinite, 
meaning that $A\geq F$ holds in the sense of matrix inequality. 
Hence together with the well-known fact that the quantum Fisher information is the 
supremum of all the induced classical Fisher information \cite{Caves 1994}, we now have 
\begin{equation}
\label{matrix ineq}
     A\geq F \geq F^C ~~ \Leftrightarrow ~~ A^{-1} \leq F^{-1} \leq (F^C)^{-1}, 
\end{equation}
where the existence of $(F^C)^{-1}$ is assumed. 
This general inequality indicates that ITE does not so much care the metric; 
hence if $F$ or $F^C$ stretches the gradient vector too much, which is problematic at 
around the target point, then switching the strategy to ITE or the ordinary gradient would 
be recommended.


\section{Example 1: single qubit}

Let us begin with the single qubit case. 
The goal is to drive the ansatz state 
\[
     \ket{\phi(\theta)}
      = \cos\theta_1\ket{0} + e^{2i\theta_2}\sin\theta_1\ket{1}
      = \left[\begin{array}{c}
            \cos\theta_1 \\
            e^{2i\theta_2}\sin\theta_1 \\
        \end{array}\right] 
\]
to the ground state of the Hamiltonian $H=\sigma_x$. 
It is clear that the north pole ($\theta_1=0$) and the south pole ($\theta_1=\pi/2$) in the 
Bloch sphere are singular points, where $\ket{\phi(\theta)}$ does not depend on $\theta_2$. 
Now the Fubini-Study metric \eqref{FS metric} is calculated as 
\begin{equation*}
   F= \left[\begin{array}{cc}
            1 & 0 \\
            0 & \sin^2(2\theta_1) \\
        \end{array}\right]. 
\end{equation*}
The two singular points are correctly characterized by the points such that ${\rm det}(F)=0$. 
On the other hand the matrix in ITE \eqref{ITE} is 
\begin{equation*}
   A= \left[\begin{array}{cc}
            1 & 0 \\
            0 & 4\sin^2(\theta_1) \\
        \end{array}\right], 
\end{equation*}
which does not capture the singularity of $\ket{\phi(\theta)}$ at $\theta_1=\pi/2$. 
The energy function is 
$f(\theta) = \bra{\phi(\theta)} H \ket{\phi(\theta)}=\sin(2\theta_1)\cos(2\theta_2)$. 
The gradient vector of $f(\theta)$ is obtained as 
\[
     \frac{\partial f(\theta)}{\partial \theta}
      = \left[\begin{array}{c}
            2\cos(2\theta_1)\cos(2\theta_2) \\
            -2\sin(2\theta_1)\sin(2\theta_2) \\
        \end{array}\right]. 
\]
All the numerical simulation shown below are based on the above analytic expressions, and no 
approximation is made. 
Also $\eta_k$ is fixed to $\eta_k=0.05$ for all $k$.

\begin{figure}[htbp]
\begin{center}
\includegraphics[width=7cm]{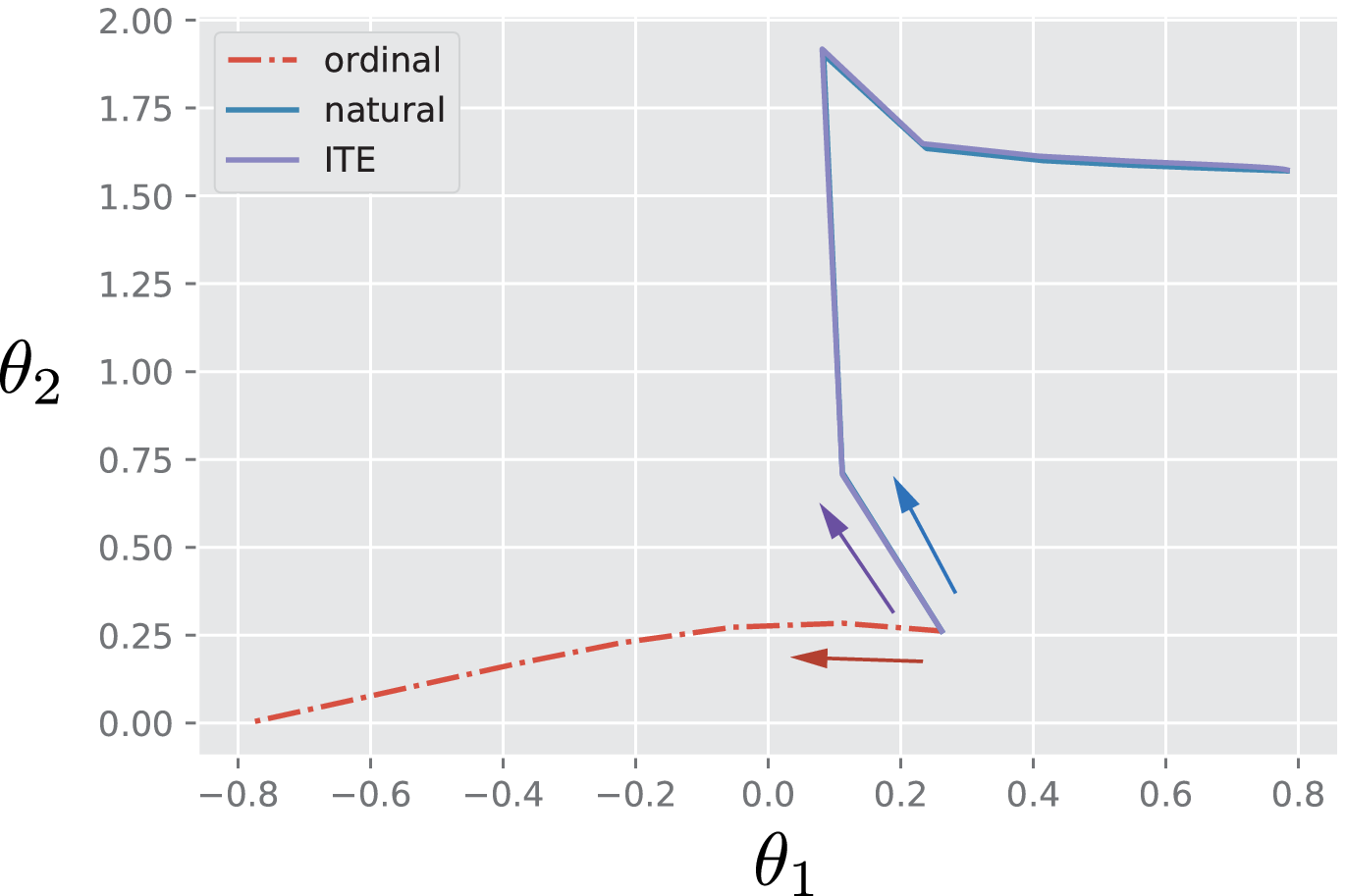}
\includegraphics[width=7cm]{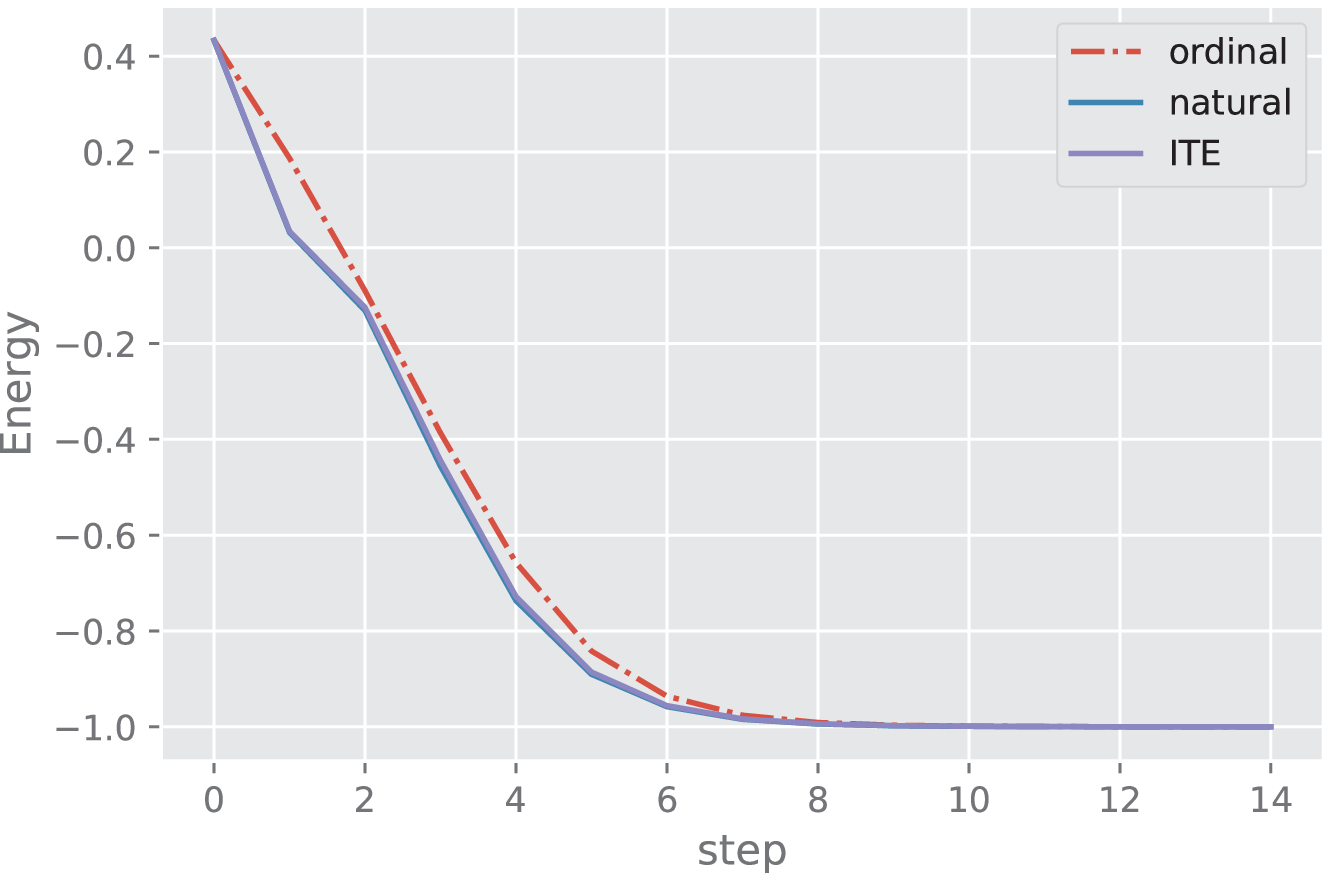}
\caption{
(Top) Trajectories of the parameters $(\theta_1, \theta_2)$ for the ordinary and natural 
gradients together with ITE, with the initial point $(\theta_1, \theta_2)=(\pi/12, \pi/12)$. 
(Bottom) Energy versus the VQE iteration steps. 
}
\label{Fig; qubit 1}
\end{center}
\end{figure}

First let us see the case where the initial point of parameters is given by 
$P_0=(\theta_1, \theta_2)=(\pi/12, \pi/12)$. 
The point of this choice is as follows; 
now $P_*=(\theta_1, \theta_2)=(-\pi/4, 0)$ is the optimum point closest to $P_0$ 
and $\tilde{P}_*=(\theta_1, \theta_2)=(\pi/4, \pi/2)$ is the second-closest optimum point 
in the Euclidean metric; 
however, now the line $\theta_1=0$ constitutes a singular point, and those distances might 
change depending on the metric because the path from $P_0$ to $P_*$ must cross this 
point while the path from $P_0$ to $\tilde{P}_*$ does not. 
In fact, as shown in Fig.~\ref{Fig; qubit 1}, the natural gradient and ITE find $\tilde{P}_*$ 
as the closest target. 
As a result, the natural gradient and ITE realize faster convergence to the ground state 
compared to the ordinary method.

\begin{figure}[htbp]
\includegraphics[width=7cm]{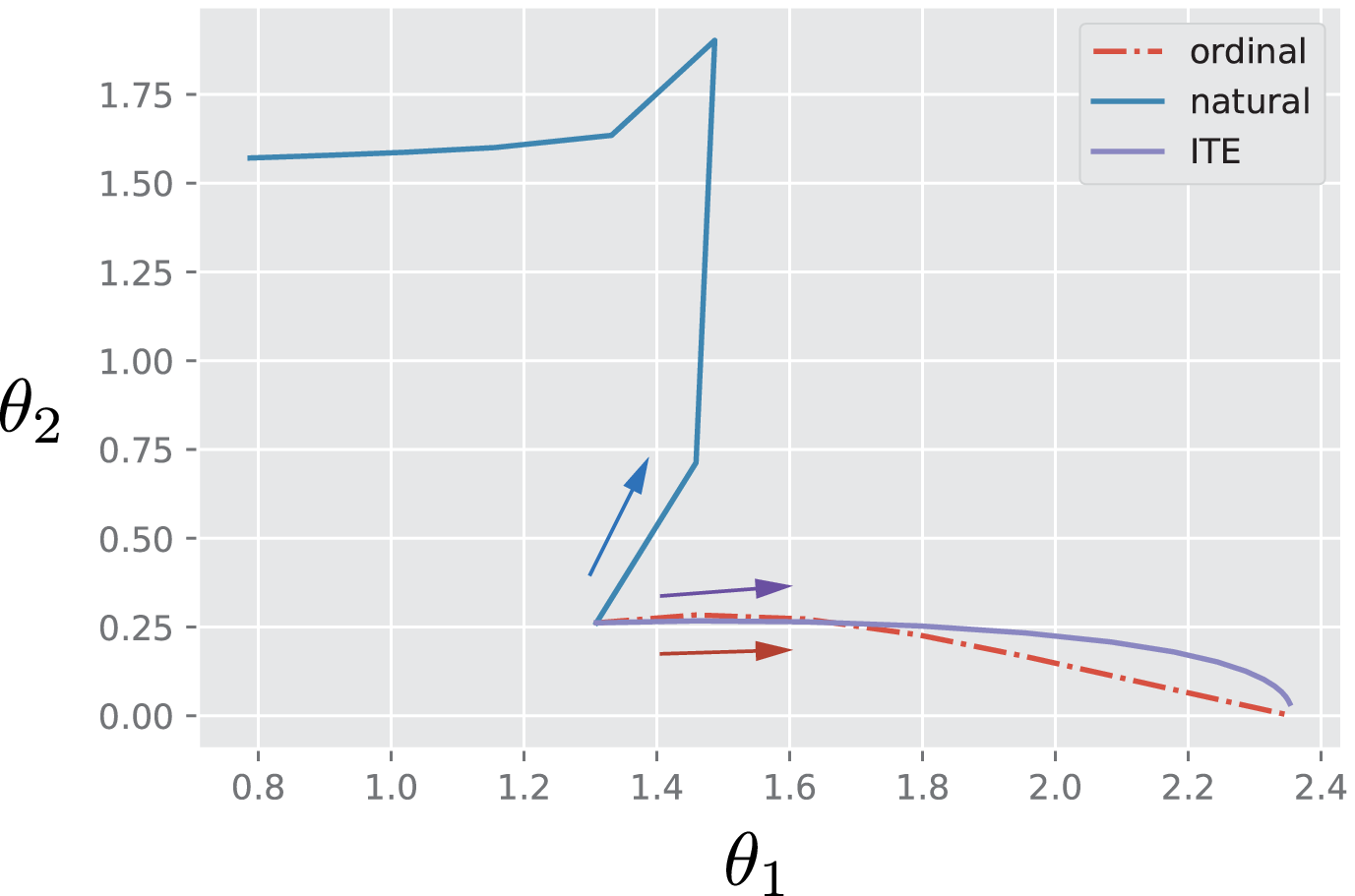}
\includegraphics[width=7cm]{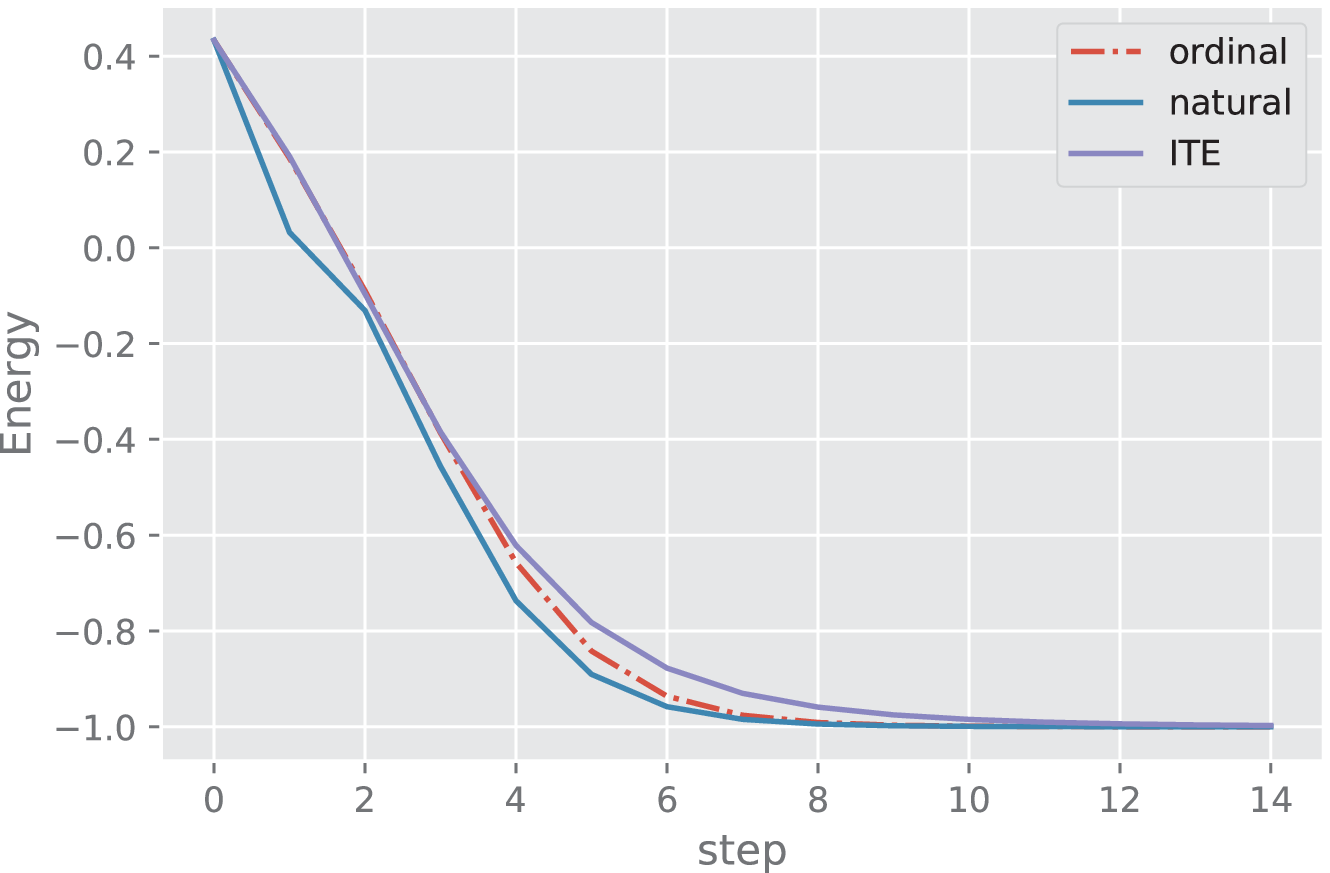}
\caption{
(Top) Trajectories of the parameters $(\theta_1, \theta_2)$ for the ordinary and natural 
gradients together with ITE, with the initial point $(\theta_1, \theta_2)=(5\pi/12, \pi/12)$. 
(Bottom) Energy versus the VQE iteration steps. 
}
\label{Fig; qubit 2}
\end{figure}

Next let us see the case where the initial point is given by 
$(\theta_1, \theta_2)=(5\pi/12, \pi/12)$. 
Note that in this case there is a singular point corresponding to $\theta_1=\pi/2$ near the 
initial point, but this singularity is recognized only by the natural gradient. 
Then, as shown in Fig.~\ref{Fig; qubit 2}, the ordinary gradient and ITE find 
$(\theta_1, \theta_2)=(3\pi/4, 0)$ as the closest optimum point, while the natural one takes 
the path toward $\tilde{P}_*$ according to the metric. 
Consequently, the natural gradient reaches the ground state first. 
Therefore we can conclude that the natural gradient actually makes use of the geometry 
of parameter space and realizes the fast convergence to the target ground state.

Lastly let us discuss the case using the classical Fisher metric \eqref{C Fisher metric} 
induced from the measurement of $H$ for the natural gradient method. 
Now the outcome is $h_{+1}=+1$ with probability 
$p_+(\theta)=|\pro{x+}{\phi(\theta)}|^2=(1+\sin(2\theta_1)\cos(2\theta_2))/2$ 
or $h_{-1}=-1$ with probability $p_-(\theta)=1 - p_+(\theta)$. 
That is, our classical probability distribution is the Bernoulli one 
$p(\theta) = \{ p_+(\theta),~p_-(\theta) \}$. 
The $2\times 2$ Fisher information matrix is then given by 
\begin{eqnarray*}
& & \hspace*{-1em}
     F^C_{ij} = p_+\frac{\partial \log p_+}{\partial \theta_i}
                         \frac{\partial \log p_+}{\partial \theta_j}
                  + p_-\frac{\partial \log p_-}{\partial \theta_i}
                          \frac{\partial \log p_-}{\partial \theta_j} 
\nonumber \\ & & \hspace*{0.7em}
     = \frac{1}{p_+ p_-}\frac{\partial p_+}{\partial \theta_i}\frac{\partial p_+}{\partial \theta_j}, 
\end{eqnarray*} 
which is clearly of rank-1, without respect to the form of $p_+(\theta_1, \theta_2)$. 
This means that the whole 2-dimensional parameter space is singular, and the natural 
gradient cannot be directly applied. 
However, because $F^C$ depends on $H$ whereas $F$ does not, the natural gradient 
for VQE with the classical Fisher information might be effectively applied to a complicated 
Hamiltonian.


\section{Example 2: ${\rm H}_2$ molecule}

\begin{figure}[htbp]
\includegraphics[width=7cm]{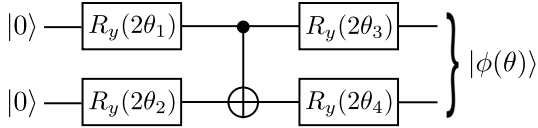}
\caption{
The hardware-efficient ansatz with 2 qubits. 
}
\label{H2 ansatz}
\end{figure}

The second case-study is on the problem of finding the ground state of the ${\rm }H_2$ 
molecule; 
the Hamiltonian can be reduced and modeled using two qubits as \cite{Temme 2017} 
\begin{equation}
\label{H2 Hamiltonian}
     H = \alpha(\sigma_z\otimes I + I \otimes \sigma_z) + \beta \sigma_x \otimes \sigma_x, 
\end{equation}
where $\alpha=0.4$ and $\beta=0.2$. 
This has four eigenvalues 
\[
    h_1 = \sqrt{4\alpha^2+\beta^2}, ~ h_2=\beta, ~ h_3=-\beta, ~ h_4 = -\sqrt{4\alpha^2+\beta^2}, 
\]
and particularly the minimum eigenvector, i.e., the ground state, is given by 
\begin{equation}
\label{H2 ground state}
    \ket{\phi_{\rm min}} \propto -\beta \ket{0, 0} + (2\alpha+\sqrt{4\alpha^2+\beta^2})\ket{1,1}. 
\end{equation}
The ansatz is taken as 
\begin{eqnarray*}
& & \hspace*{-1em}
     \ket{\phi(\theta)} 
\nonumber \\ & & \hspace*{-0.8em}
       = (R_y(2\theta_3) \otimes R_y(2\theta_4)) U_{\rm ent} 
                                       (R_y(2\theta_1) \otimes R_y(2\theta_2)) \ket{0}\otimes\ket{0},
\end{eqnarray*} 
where $U_{\rm ent}=\ket{0}\bra{0}\otimes I + \ket{1}\bra{1}\otimes \sigma_x$ denotes 
the CNOT gate and $R_y(\theta)$ denotes the single-qubit rotation operator defined by 
\[
          R_y(\theta) = e^{-i\theta \sigma_y/2} 
           = \left[\begin{array}{cc}
                  \cos(\theta/2) & -\sin(\theta/2) \\
                  \sin(\theta/2) & \cos(\theta/2) \\
              \end{array}\right]. 
\]
This is a typical hardware-efficient ansatz, illustrated in Fig.~\ref{H2 ansatz}.

The Fubini-Study metric \eqref{FS metric} is calculated as 
\begin{equation*}
   F  = \left[\begin{array}{cc|cc}
              1 & 0 & \sin(2\theta_2) & 0 \\
              0 & 1 & 0 & \cos(2\theta_1) \\ \hline 
              \sin(2\theta_2) & 0 & 1 & 0 \\
              0 & \cos(2\theta_1) & 0 & 1\\
           \end{array}\right]. 
\end{equation*}
Note that $\ket{\phi(\theta)}$ is a real vector and thus $\pro{\partial_i \phi}{\phi}=0$ 
for all $i$. 
Hence $F=A$; that is, ITE is equivalent to the natural gradient. 
Now the determinant of $F$ is given by 
\[
    {\rm det}(F) = \sin^2(2\theta_1)\cos^2(2\theta_2). 
\]
The parameters satisfying ${\rm det}(F)=0$ constitute the set of singular points, which 
has a clear physical meaning as follows. 
In general, the entanglement of the bipartite state $\ket{\Psi}$ can be quantified by 
the entanglement entropy 
\[
     S(\ket{\Psi}) = -{\rm Tr}(\rho_1\log\rho_1), ~~
     \rho_1 ={\rm Tr}_2(\ket{\Psi}\bra{\Psi}).
\]
In our case, it is given by 
\[
    S(\ket{\phi}) = -\lambda\log\lambda - (1-\lambda)\log(1-\lambda), 
\]
where 
\[
    \lambda = \frac{1}{2} + \frac{1}{2}\sqrt{1-{\rm det}(F)}. 
\]
Hence, $S(\ket{\phi})=0$ if and only if ${\rm det}(F)=0$. 
That is, the set of singular points represents the set of all separable states. 
This makes sense, because if the state 
$U_{\rm ent}(R_y(2\theta_1) \otimes R_y(2\theta_2)) \ket{0}\otimes\ket{0}$ is separable, 
then the local operation $R_y(2\theta_3) \otimes R_y(2\theta_4)$ can never entangle 
this state for any parameter choice. 

\begin{figure}[htbp]
\includegraphics[width=8cm]{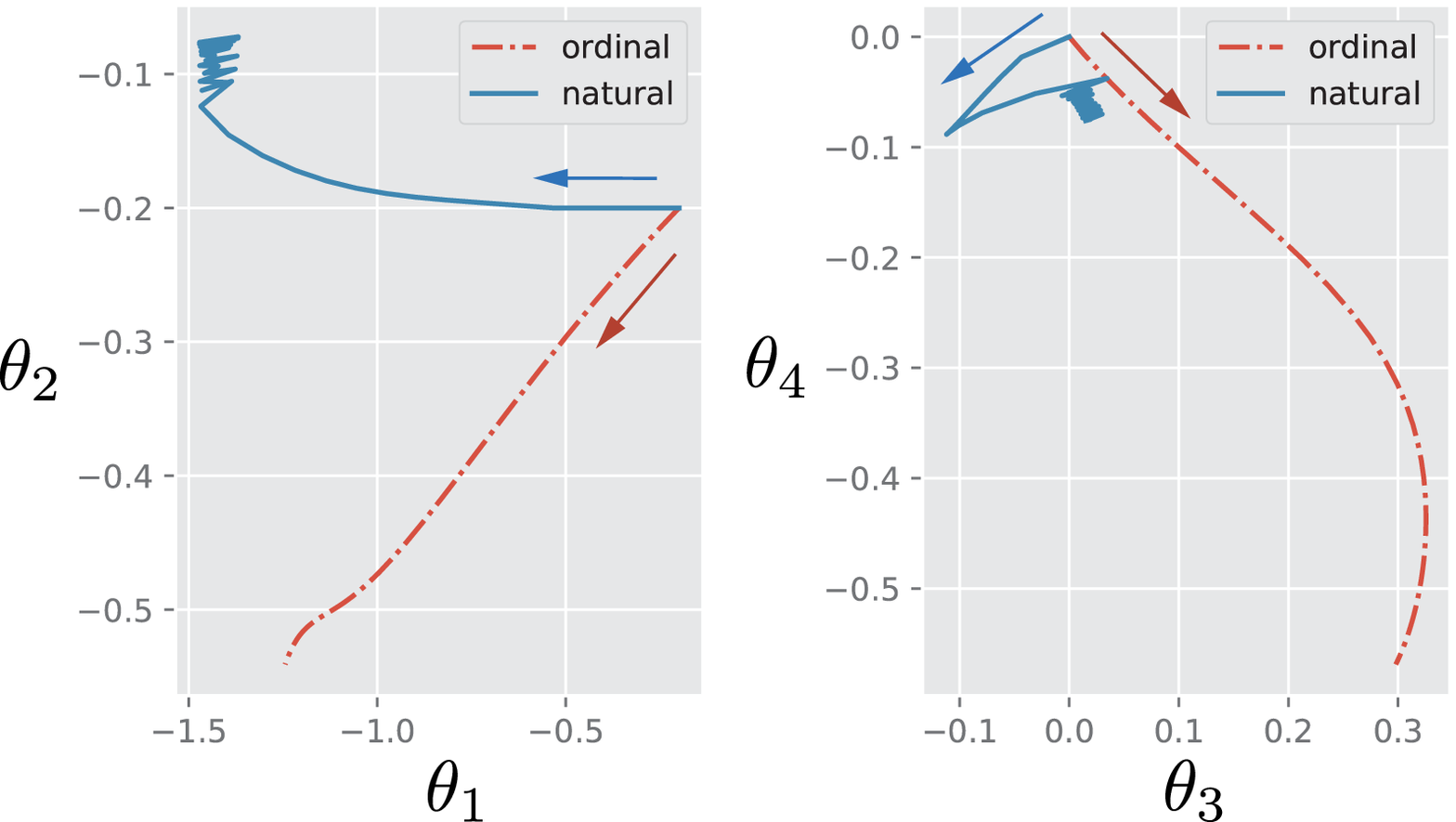}
\includegraphics[width=7cm]{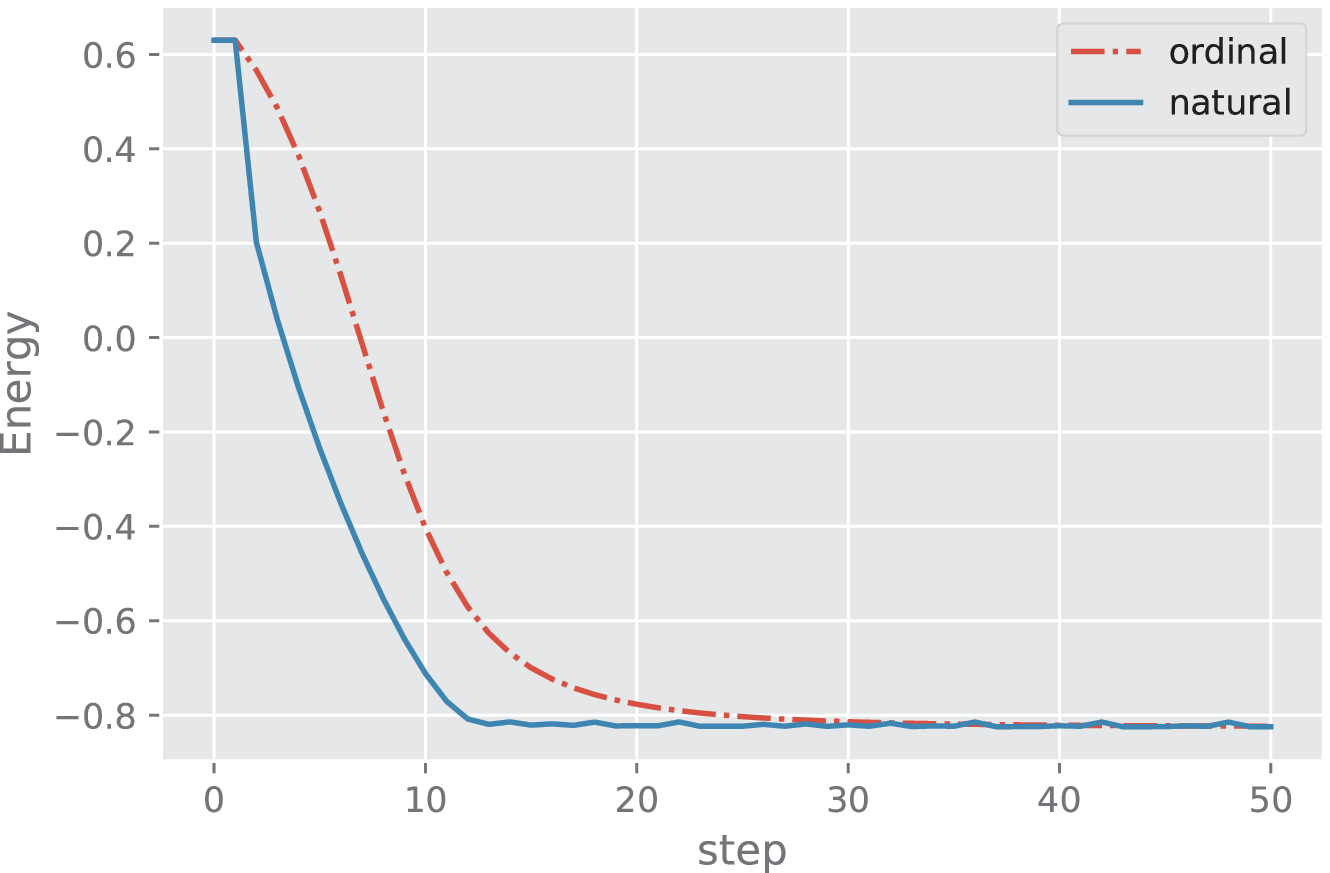}
\caption{
(Top) Trajectories of the parameters $(\theta_1, \theta_2, \theta_3, \theta_4)$ for the 
ordinary and natural gradients, with the initial point 
$(\theta_1, \theta_2, \theta_3, \theta_4)=(-0.2, -0.2, 0, 0)$. 
(Bottom) Energy of the ${\rm H}_2$ molecule ($\alpha=0.4, \beta=0.2$) versus the 
VQE iteration steps. 
}
\label{Fig; H2 1}
\end{figure}

Let us now see the results of numerical simulation. 
Again the learning coefficient is fixed to $\eta_k=0.05$ for all $k$. 
First, Fig.~\ref{Fig; H2 1} shows the trajectory of the parameter dynamics (Top) and the 
change of $f(\theta)$ over the VQE iteration step, for the case where the initial point is 
$(\theta_1, \theta_2, \theta_3, \theta_4)=(-0.2, -0.2, 0, 0)$. 
As shown in the figure, the natural gradient achieves the faster convergence to the 
minimum energy $h_4\approx -0.82$ than the ordinary gradient. 
This faster convergence might be explained as follows. 
If the initial value of $\theta_3$ and $\theta_4$ are chosen as $(\theta_3, \theta_4)=(0, 0)$, 
the initial state is 
\begin{equation}
\label{H2 initial state}
     \ket{\phi(\theta)} 
       = \left[\begin{array}{c}
             \cos\theta_1\cos\theta_2 \\
             \cos\theta_1\sin\theta_2 \\
             \sin\theta_1\sin\theta_2 \\
             \sin\theta_1\cos\theta_2 \\
          \end{array}\right]. 
\end{equation}
Hence, if $\theta_2$ is nearly zero, the initial state is already close to the target ground state 
\eqref{H2 ground state}, meaning that $(\theta_2, \theta_3, \theta_4)$ need not be largely 
changed. 
It seems that the natural gradient effectively utilizes this fact, as seen in Fig.~\ref{Fig; H2 1}.

\begin{figure}[htbp]
\includegraphics[width=7cm]{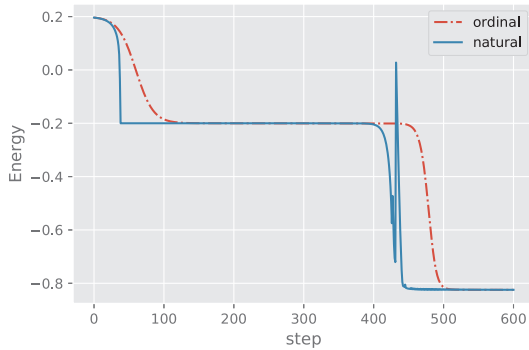}
\caption{
Energy of the H2 molecule ($\alpha=0.4, \beta=0.2$) versus the VQE iteration steps, 
with the initial point $(\theta_1, \theta_2, \theta_3, \theta_4)=(7\pi/32, \pi/2, 0, 0)$. 
}
\label{Fig; H2 2}
\end{figure}

Next let us see the case where the initial value of parameters are 
$(\theta_2, \theta_3, \theta_4)=(\pi/2, 0, 0)$; 
in this case, the initial state \eqref{H2 initial state} is 
$\ket{\phi}=[0, \cos\theta_1, \sin\theta_1, 0]^\top$, which is thus close to the first or second 
excited state $(\ket{0,1}\pm\ket{1,0})/\sqrt{2}$. 
Hence the parameters have to be largely changed to transfer the initial state to the 
target ground state. 
Then there might be some singular points along this passage; 
the natural gradient may have an advantage in such a case, by stretching the gradient 
vector depending on the metric, as explained in Fig.~\ref{Grad Flow image}. 
Figure~\ref{Fig; H2 2} plots the energy with the initial value $\theta_1=7\pi/32$, which 
indeed demonstrates this desirable scenario; 
that is, at around the first excited state with energy $-\beta=-0.2$, the natural gradient 
can efficiently search the path to get out the plateau and move toward the ground state 
faster than the ordinary gradient. 
Note that during the second drop-off, the natural gradient experiences a steep rise in 
energy, meaning that actually the path in the parameter space moves near a singular 
point such that $F^{-1}$ takes a large value. 
Such a sudden change of the parameters can be avoided by adding a small positive 
number to the eigenvalues of $F$ via the singular value decomposition of $F$.

\begin{figure}[htbp]
\includegraphics[width=7cm]{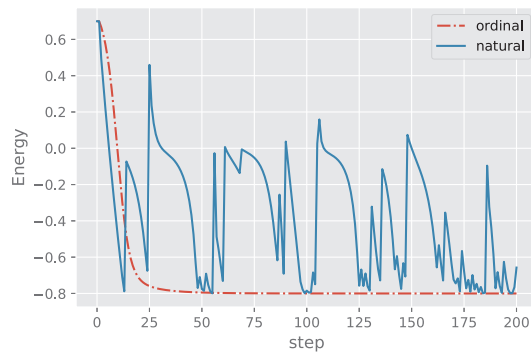}
\caption{
Energy of the toy molecule ($\alpha=0.4, \beta=0.02$) versus the VQE iteration steps, 
with the initial point $(\theta_1, \theta_2, \theta_3, \theta_4)=(-0.2, -0.2, 0, 0)$. 
}
\label{Fig; H2 3}
\end{figure}

Lastly we discuss the toy molecule having Hamiltonian \eqref{H2 Hamiltonian} with 
$\alpha=0.4$ and $\beta=0.02$. 
In this case, the target ground state \eqref{H2 ground state} is close to the separable state 
$\ket{1,1}$ with minimum energy $h_4\approx-2\alpha=-0.8$. 
This should be problematic for the natural gradient, because, as seen above, all the 
separable states correspond to the singular points in the parameter space; 
as a result, in this case the natural gradient vector must be largely stretched near the target 
ground state. 
This is actually seen in Fig.~\ref{Fig; H2 3}, showing that the state with natural gradient never 
stay at the ground state. 
In such a case the learning coefficient $\eta_k$ should be modified to monotonically decrease, 
e.g., $\eta_k=1/k$; but this strategy did not work well for this problem. 
Hence what we have learned from this case-study on the use of natural gradient is that 
the metric should be carefully analyzed so that the target state (which is however unknown) 
would not lie near singular points in the parameter space.

\section{Conclusion}

Some case-studies on the natural gradient for the VQE problems have been discussed 
in this paper. 
It is hoped that the reader would gain some insight about how to make use of the 
geometric property of a parametrized ansatz state to effectively apply the natural gradient 
method, possibly even to a relatively large VQE problem.

This work was supported by MEXT Quantum Leap Flagship Program Grant Number 
JPMXS0118067285. 
The author acknowledges helpful discussion with Y. Onishi and Y. Yamaji.





\begin{thebibliography}{}



\bibitem{O'Brien 2014}
A. Peruzzo, J. McClean, P. Shadbolt, M. Yung, X. Zhou, P. J. Love, A. Aspuru-Guzik, and J. L. O'Brien, 
A variational eigenvalue solver on a photonic quantum processor, 
Nat. Commun. 5, 4213 (2014).

\bibitem{McClean 2016}
J. R. McClean, J. Romero, R. Babbush, and A. Aspuru-Guzik, 
The theory of variational hybrid quantum-classical algorithms, 
New J. Phys. 18, 023023 (2016). 

\bibitem{Kandala 2017}
A. Kandala, A. Mezzacapo, K. Temme, M. Takita, M. Brink, J. M. Chow, and J. M. Gambetta, 
Hardware-efficient variational quantum eigensolver for small molecules and quantum magnets, 
Nature 549, 242 (2017). 

\bibitem{Temme 2017}
S. Bravyi, J. M. Gambetta, A. Mezzacapo, and K. Temme, 
Tapering off qubits to simulate fermionic Hamiltonians, 
arXiv:1701.08213 (2017). 

\bibitem{AdptVQE 2019}
H. R. Grimsley, S. E. Economou, E. Barnes, and N. J. Mayhall, 
An adaptive variational algorithm for exact molecular simulations on a quantum computer, 
Nat. Commun. 10, 3007 (2019).

\bibitem{Fujii 2019}
N. Yoshioka, Y. O. Nakagawa, K. Mitarai, and K. Fujii, 
Variational quantum algorithm for non-equilibrium steady states, 
arXiv:1908.09836 (2019).



\bibitem{SPSA}
J. C. Spall, 
Multivariate stochastic approximation using a simultaneous perturbation gradient approximation, 
IEEE Trans. Autom. Contr. 37, 332-341 (1992). 




\bibitem{Amari 1998}
S. Amari, 
Natural gradient works efficiently in learning, 
Neural Computation, 10, 251/276 (1998). 




\bibitem{Carleo 2017}
G. Carleo and M. Troyer, 
Solving the quantum many-body problem with artificial neural networks, 
Science 355, 602 (2017). 

\bibitem{Cirac 2018}
I. Glasser, N. Pancotti, M. August, I. D. Rodriguez, and J. I. Cirac, 
Neural-network quantum states, string-bond states, and Chiral topological states, 
Phys. Rev. X 8, 011006 (2018).



\bibitem{Sorella 2001}
S. Sorella. 
Generalized Lanczos algorithm for variational quantum Monte Carlo, 
Phys. Rev. B 64, 024512 (2001). 

\bibitem{Sorella 2004}
M. Casula C. Attaccalite, and S. Sorella. 
Correlated geminal wave function for molecules: An efficient resonating valence bond approach. 
Journal. Chem. Phys. 121, 7110 (2004). 


\bibitem{Benjamin 2018a}
S. McArdle, T. Jones, S. Endo, Y. Li, S. Benjamin, and X. Yuan, 
Variational quantum simulation of imaginary time evolution, 
arXiv:1804.03023 (2018).

\bibitem{Benjamin 2018b}
X. Yuan, S. Endo, Q. Zhao, S. Benjamin, and Y. Li, 
Theory of variational quantum simulation, 
arXiv:1812.08767 (2018).



\bibitem{Carleo 2019}
J. Stokes, J. Izaac, N. Killoran, and G. Carleo, 
Quantum natural gradient, 
arXiv:1909.02108 (2019). 

\bibitem{Caves 1994}
S. L. Braunstein and C. M. Caves, 
Statistical distance and the geometry of quantum states, 
Phys. Rev. Lett. 72, 3439 (1994). 


\end{thebibliography}
\end{document}